\documentclass[prb,twocolumn,showpacs,amsmath,amssymb,superscriptaddress]{revtex4}

\usepackage{graphicx}
\usepackage{dcolumn}
\usepackage{bm}
\usepackage{color}
\usepackage{amssymb}

\newcounter{mycounter} 
\begin{document}

\title{Quantized excitation spectra by magnon confinement in quasi-one-dimensional S=1 spin systems}

\author{Takafumi Suzuki }
\author{ Sei-ichiro Suga}

\affiliation{Graduate School of Engineering, University of Hyogo, Himeji 671-2280, Japan}

\date{\today}

\begin{abstract} 
The infinite time-evolving block decimation algorithm is applied to calculate the dynamical spin-structure factors of the quasi-one-dimensional (Q1D) S=1 antiferromagnetic spin system with single-ion anisotropy and bond alternation.
It is found that when the staggered field induced by the weak interchain interaction is taken into account, the excitation continuum originating from magnons is quantized.
The excitation energies of the quantized excitation spectra are well-explained by the negative zeros of the Airy functions, when the single-ion anisotropy is negatively strong.
This quantization of the magnon continuum is a counterpart of the spinon confinement, which has been recently discussed for Q1D S=1/2 antiferromagnets. 
It is further shown that, when a staggered field exists, the quantized excitation spectra appear on the phase boundary between the Haldane and N\'eel phases of the phase diagram without the staggered field.
However, the quantized excitation spectra disappear in the singlet-dimer phase. 
\end{abstract}

\preprint{APS/123-QED}

\maketitle

Recent inelastic neutron scattering (INS) experiments\cite{BGrenier,QFaure} on quasi-one-dimensional (Q1D) S=1/2 antiferromagnetic (AF) Ising-like XXZ magnets, (Ba/Sr)Co$_2$V$_2$O$_8$\cite{SKimura}, 
 have reported that, below the N\'eel temperature $T_{\rm N}$, these materials show quantized excitation spectra with excitation energies that are well explained by the series of negative zeros in the Airy function (NZAF).
McCoy and Wu argued the relation between the NZAF and confinement of the domain-wall excitations by focusing on the excitation spectra of the S=1/2 ferromagnetic (FM) Ising chain in weak transverse and longitudinal magnetic fields\cite{McCoy,SM}.
Since the ground state of this model is a fully polarized state, low-energy excitation is achieved by flipping spins of arbitrary lengths, i.e., two domain-wall excitations.
Since each domain wall carries $\Delta S=1/2$, the low-energy excitation is interpreted as two-spinon excitation.
In the transverse field, these two spinons travel in the chain and compose an excitation continuum.
When the uniform longitudinal field is further applied weakly, it works as a linear potential between the two spinons.
The effective model for this two-spinon excitation is exactly solvable and the eigenvalues are given by NZAF~\cite{McCoy,SM}.
This means that the excitation spectra of the two-spinon continuum are quantized.
The quantized spectra explained by NZAF have been confirmed in the INS experiments\cite{RColdea} on ${\rm CoNb_2O_6}$, which is a ferromagnetic Ising-spin-chain compound.
Shiba also discussed the quantized excitation spectra in the Q1D S=1/2 AF Ising-like XXZ system\cite{HShiba}.
When we focus on bipartite systems, the interchain interaction in the Q1D AF systems effectively works as a weak staggered field below $T_{\rm N}$. 
Thus, it is expected that the domain-wall excitations along the spin-chain direction are confined by the weak staggered field and that the same discussion as that for the ferromagnetic Ising-spin chain is applicable for low-energy excitation.

The two-spinon confinement plays a key role in the quantized excitation spectra.
On the other hand, the excitation continuum can be generated by other quasi-particles.   
A multi-magnon continuum in the S=1 AF Heisenberg chain where a single magnon carries $\Delta S=1$ is a typical candidate.
In the S=1 AF Heisenberg chain, the ground state is the celebrated Haldane-gap state\cite{Haldane1,Haldane2,AKLT}.
It has been shown that, for low-energy excitation, the single-magnon isolated mode appears at $q \approx \pi$, while the lower edge of the two-magnon continuum is expected to be $q \approx 0$\cite{SRWhite2,IAffleck,SYamamoto,MDPHorton,MTakahashi,FHLEssler}.
At $q \approx \pi$, the lower edge of the three magnon continuum appears above the single magnon mode\cite{SRWhite2,IAffleck,SYamamoto,MDPHorton,MTakahashi,FHLEssler}.
Despite these studies, however, low-energy excitations in the Q1D S=1 AF spin system below $T_{\rm N}$ are still poorly understood.
In this Rapid Communication, we focus on the quantized spectra in the Q1D S=1 AF spin system with single-ion anisotropy and bond alternation.

%
%
%
%
%
%
The Hamiltonian for Q1D S=1 AF spin system with single-ion anisotropy and bond alternation is written as
\begin{eqnarray}
{\mathcal H} =& J  {\displaystyle \sum_{ij}} (1+\alpha(-1)^{i} ) {\boldsymbol S}_{i,j} \cdot {\boldsymbol S}_{i+1,j}\nonumber \\
&+ D {\displaystyle \sum_{ij}} {S_{i,j}^z}^2+J' {\displaystyle \sum_{i \langle j,j' \rangle}} {\boldsymbol S}_{i,j} \cdot {\boldsymbol S}_{i,j'}, 
\end{eqnarray}
where $i$ is the site index in the chain direction and $j$ specifies the chain.
$J$($>$ 0) is antiferromagnetic intra-chain interaction and $J'$ ($>$ 0) is the interchain interaction. 
$\alpha$ denotes the bond alternation in the spin-chain direction and $D$ represents the single-ion anisotropy.
The summation $\langle j,j' \rangle$ runs over all nearest-neighbor interchain pairs.
The bipartite system is assumed.

When the mean-field treatment for the weak interchain interaction is applied, the staggered fields are induced in the intra-chain Hamiltonian.
The effective Hamiltonian reads
\begin{eqnarray}
\mathcal{H}_{\rm MF}= &J {\displaystyle \sum_{i}} (1+\alpha(-1)^{i}){\boldsymbol S}_{i} \cdot {\boldsymbol S}_{i+1}\nonumber \\
&  + D {\displaystyle \sum_{i}}  {S_{i}^z}^2  + h_s {\displaystyle \sum_{i}} (-1)^{i}{S_{i}}^z,
\label{Ham2}
\end{eqnarray}
where $h_s$ is the mean field derived from the magnetic moments of the nearest-neighbor chains and is determined by solving a self-consistent equation.
In the following calculations, we set a small value of $h_s$, because it is not important to obtain $h_s$ self-consistently from a given $J'$.

We apply the infinite time-evolving block decimation algorithm\cite{GVidal,ROrus} to calculate the dynamical spin structure factor (DSF) for the Hamiltonian (\ref{Ham2}). 
The DFS is defined as $S^{\mu\mu}(q,\omega)=\pi^{-1}{\rm Im} \int \int i A^{\mu\mu}(x,t) e^{-iqx-i(\omega-e_g)t}dxdt$, 
where $A^{\mu\mu}(x,t)=\langle {S_x}^{\mu}(t){S_0}^{\mu}(0)\rangle$ is the dynamical spin-spin correlation function and $e_g$ is the ground-state energy. Note that ${S_x}^{\mu}(t)=e^{i{\mathcal H}_{\rm MF}t}{S_x}^{\mu}e^{-i{\mathcal H}_{\rm MF}t}$.  
The details of the numerical techniques have been discussed in Ref. \onlinecite{HNPhien}.
We obtain the DSFs from the Fourier transformation of $A^{\mu\mu}(x,t)$ for a finite window size $N$ in real space.
To reduce numerical noise, we combine the Gaussian filtering method~\cite{SRWhite} with the Fourier transformation.
In the following calculations, we set  $\chi_{\rm max}=120$, where $\chi$ corresponds to the maximum value of the bond dimension for tensors composing the wave function, and $N=200$ for the window size.
We calculate the DSFs for S=1/2 and S=1 Heisenberg chain as a benchmark test, which are shown in Ref. \onlinecite{SM}.

\begin{figure}[htb]
\begin{center}
\includegraphics[bb=0 0 1112 1796, scale=0.18]{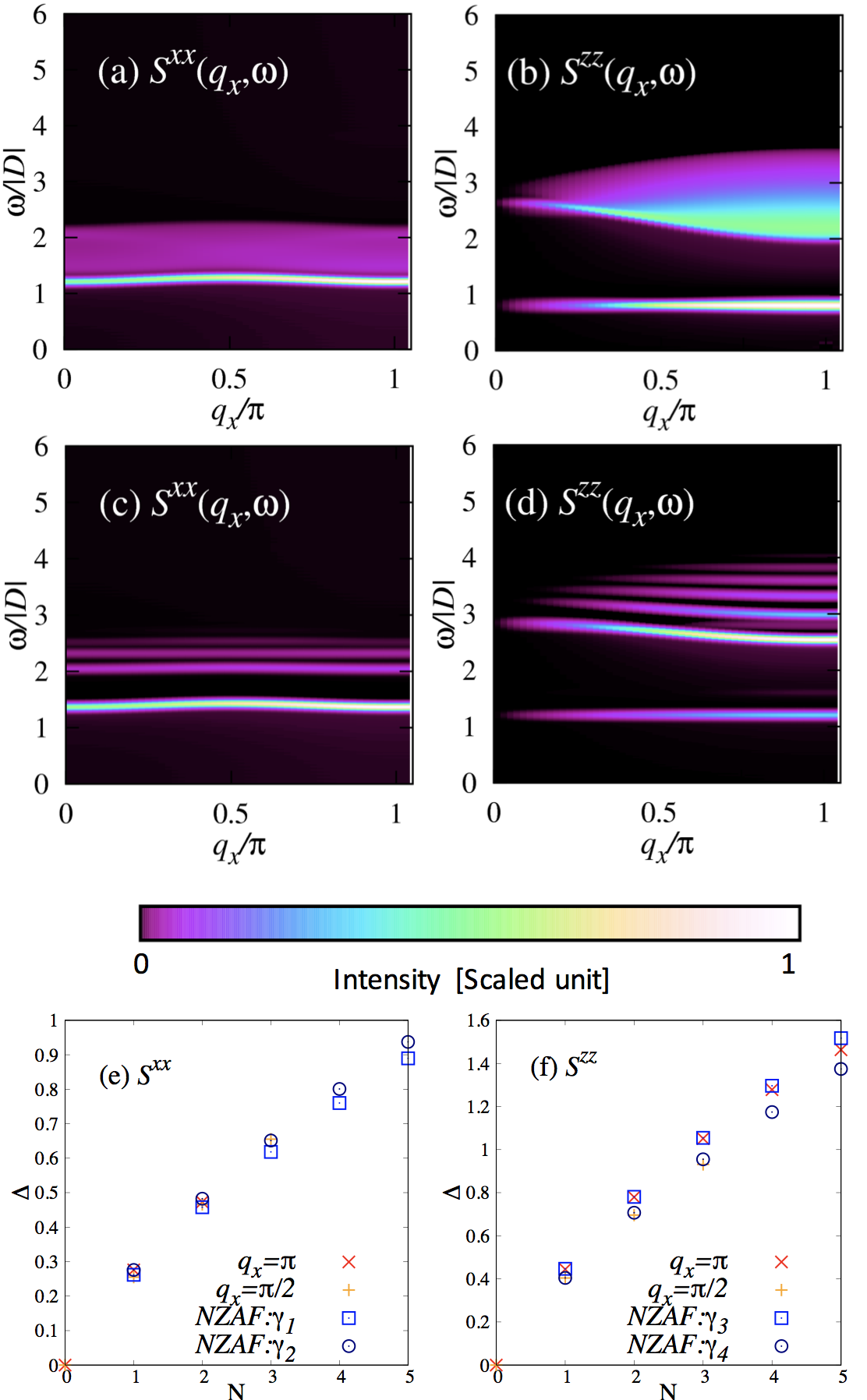}
\hspace{0pc}
\vspace{-0.5pc}
\caption{\label{fig2} $S^{xx}(q_x,\omega)$ and $S^{zz}(q_x,\omega)$ for $(\alpha,D/J)=(0,-5)$ at (a), (b) $h_s=0$ and at (c), (d) $h_s/|D|=0.1$.
The excitation energies $\Delta$ of the quantized excitation spectra of (e) $S^{xx}(q_x,\omega)$ and (f) $S^{zz}(q_x,\omega)$ at $q_x=\pi$ and $\pi/2$.
The horizontal axis corresponds to the excitation level of the quantized spectra.
Note that the isolated mode that already emerged at $h_s=0$ is excluded for counting the excitation levels.
$\Delta$ is measured from the lowest quantized energy. 
The NZAF results are scaled using a constant factor $\gamma_i$ to fit the excitation energy $\Delta(N=1)$  for the first excited state in the quantized excitation spectra, $\Delta=\gamma_i z_i$, 
where $z_i$ indicates each negative zero of the Airy function. 
The constants used for the fitting are $\gamma_1 \approx 0.270$, $\gamma_2 \approx 0.285$, $\gamma_3 \approx 0.191$, and $\gamma_4 \approx 0.173$.}
\end{center}
\end{figure}

In Fig. \ref{fig2}, we show the DSFs of the Hamiltonian (\ref{Ham2}) for $(\alpha,D/J)=(0,-5)$.
When $(\alpha,D/J)=(0,-5)$ and $h_s=0$, the ground state is the N\'eel state.
Since the critical point between the Haldane and N\'eel phases exists at $D_c/J=-0.36\pm0.01$~\cite{TTonegawa} for $\alpha=0$ and $h_s=0$, 
$(\alpha,D/J)=(0,-5)$ is considered to be located deep in the N\'eel phase.
For $(\alpha,D/J)=(0,-5)$, an isolated mode appears in $0 \le q_x \le \pi$ in the low-lying excitation [Figs. \ref{fig2}(a) and \ref{fig2}(b)].
The excitation continuum appears above the isolated mode.
These excitation continua are quantized by the finite value of $h_s$, as shown in Figs. \ref{fig2}(c) and \ref{fig2}(d).
In Figs. \ref{fig2}(e) and \ref{fig2}(f), the NZAF and each excitation energy of the quantized excitation spectra at $q_x=\pi$ and $\pi/2$ are compared.
We find that the excitation energies of the quantized spectra in both $S^{xx}(q_x,\omega)$ [$=S^{yy}(q_x,\omega)$] and $S^{zz}(q_x,\omega)$ are quantitatively explained by NZAF.
Note that the lowest quantized state is $not$ the low-lying isolated mode.


Since each spin prefers the $S^z=\pm 1$ state for the negatively large $D/J$, the system is mapped to the AF Ising chain at the Ising limit, $D/J \rightarrow -\infty$, due to the positive $J$. 
For the bipartite system, the AF Ising chain is equivalent to the FM Ising chain via spin rotations on one of the two sub-lattices.
The mapped Hamiltonian is expressed as
${\mathcal H_{\rm FM}}={\mathcal H}_0 + {\mathcal H}_1$, where ${\mathcal H}_0=D{\sum_{i}} {S^z_i}^2 - J\sum_{i}S^z_iS^z_{i+1}-h_s\sum_{i}S^z_i$ and ${\mathcal H}_1=J {\sum_{i}} \left( S^{+}_iS^{+}_{i+1} + S^{-}_iS^{-}_{i+1} \right)$.
%
%
The condition $|D|/J \gg 1$ allows ${\mathcal H}_1$ to be treated as the perturbation.
If $h_s$ is positively infinitesimal, the ground state of ${\mathcal H}_0$ is the fully polarized state expressed by $\psi^{+}_{\rm GS} = |\cdots $+++++++$ \cdots \rangle$, where $+$, $0$, and $-$ in the ket denote $S^z=1$, $0$, and $-1$, respectively.
When ${\mathcal H}_1$ is included, the total $S_z$ is not a good index, but the Hilbert space is still classified by the parity of total $S_z$.

First, we consider the low-energy excitation in $S^{xx}(q_x,\omega)$.
The low-energy excitation in $S^{xx}(q_x,\omega)$ is described by the dynamics of the state $\psi^{+}_1$ whose initial state is prepared by $\psi^{+}_1=S_i^{x}|\psi^{+}_{\rm GS}\rangle$.
Since $S_i^{x} \propto (S_i^{+}+S_i^{-})$, this initial state is interpreted as one magnon state, $\psi^{+}_1 \propto |\cdots $+++0+++$ \cdots \rangle$.
The parity of $\psi^{+}_1$ is different from that of the ground state.
Since the energy cost to create $\psi^{+}_1$ is approximately $\omega = |D|+2J$,  the isolated mode by $\psi^{+}_1$ appears at $\omega = |D|+2J$, in the absence of ${\mathcal H}_1$.
When ${\mathcal H}_1$ acts on the site with $S^z=0$, the further excited state appears above $\psi^{+}_1$: $S^z=0$ moves to the nearest-neighbor site, accompanying the site with $S^z=-1$, $| \cdots $+++$-$0++$ \cdots \rangle$.
This means that the domain composed of the $S^z=-1$ sites develops by the action of ${\mathcal H}_1$ on the site with $S^z=0$ repeatedly.
Therefore, the excitation continuum with the band center at $\omega \approx |D| + 4J$ appears above the excitation mode by the $\psi^{+}_1$ state.
In the mapped Hamiltonian, $h_s$ works as the confinement potential for the domain with $S^z=-1$ sites, and thus, the excitation continuum is quantized by $h_s$.

Next, the low-energy excitation in $S^{zz}(q_x,\omega)$ is considered. 
Since the initial state in the longitudinal component is given by multiplying the ground state at the site $i$ by $S^z_i$,
the parity of total $S^z_i$ for the initial state is conserved.
The ${\mathcal H}_1$ operation on a nearest-neighbor pair of the initial state creates the two-magnon state, namely, $\psi^{+}_2 = |\cdots $+++00+++$ \cdots \rangle$.
The energy cost for creating $\psi^{+}_2$ is $\omega \approx 2|D|+3J$.
For example, when ${\mathcal H}_1$ acts on $\psi^{+}_2$ twice at the sites with $S^{z}=0$, each magnon moves in the opposite direction, $| \cdots $++0$-$$-$0++$ \cdots \rangle$.
Thus, the domain with $S^{z}=-1$ sites develops through the action of ${\mathcal H}_1$ repeatedly on the site with $S^{z}=0$.
This means that the domain-wall excitation is also allowed for $\psi^{+}_2$ and the excitation continuum appears centered at $\omega \approx 2|D| + 4J$.
In the same manner as the excitation continuum by $\psi^{+}_1$, the excitation continuum by $\psi^{+}_2$ is quantized by $h_s$.

The quantization of the excitation continua discussed above is explained by NZAF, because the mechanism, namely, the confinement of the domain walls, is similar to that discussed for the S=1/2 FM Ising spin chain\cite{McCoy}.
The above scenario is considered to be satisfied for $D/J=-5$, which implies that the system is near the Ising limit.
From the analogy to the domain wall excitation in the S=1/2 FM Ising spin chain\cite{McCoy}, domain wall excitation by $\psi^{+}_0 = |\cdots $+++$-$$-$+++$ \cdots \rangle$ is expected to appear in the low-energy excitation. 
However, a higher-order process is required to create such a state and the intensity is much suppressed in the present case. 

\begin{figure}[htb]
\begin{center}
\includegraphics[bb=0 0 1156 1808, scale=0.18]{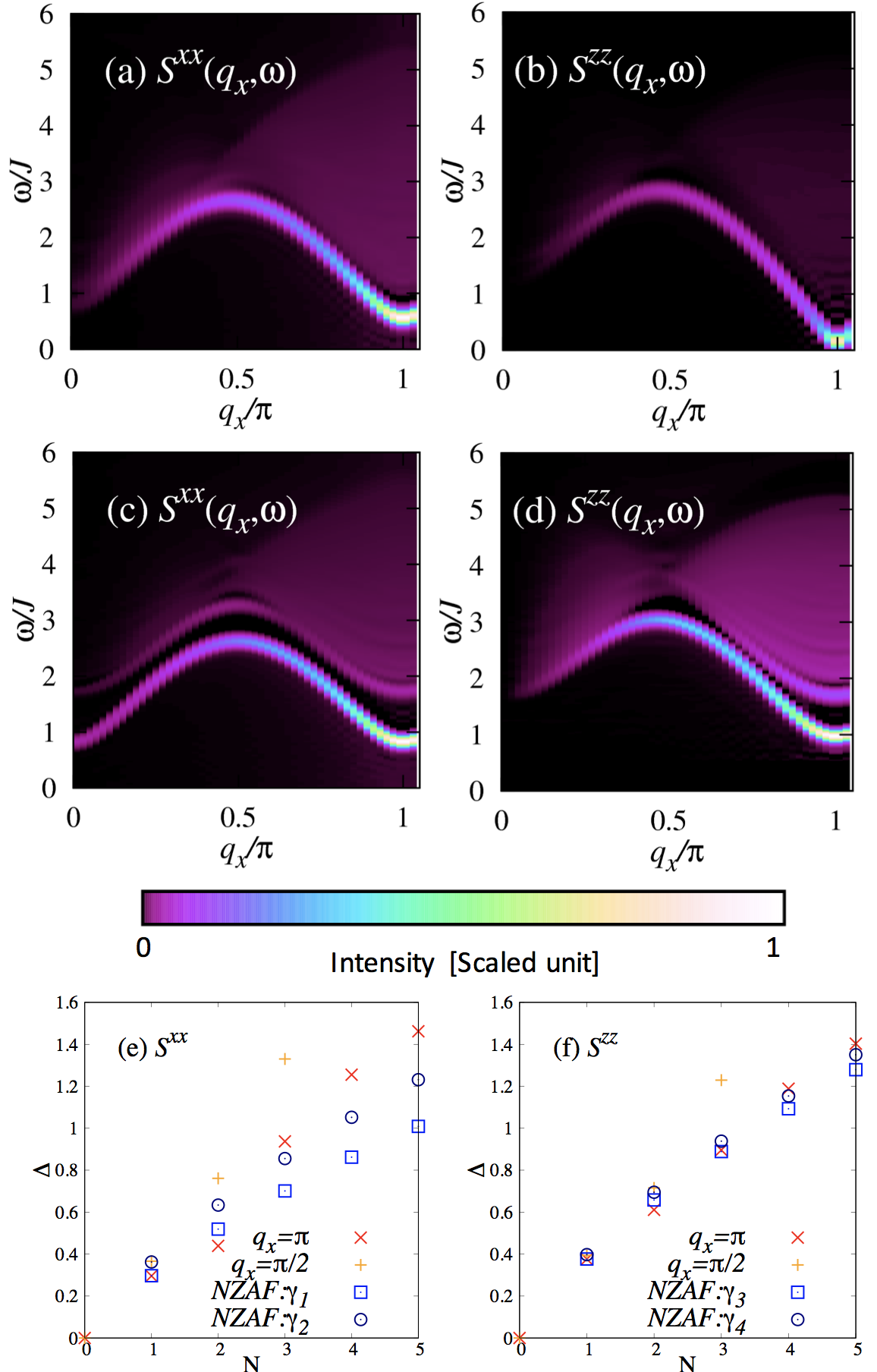}
\hspace{0pc}
\vspace{-0.5pc}
\caption{\label{fig3} $S^{xx}(q_x,\omega)$ and $S^{zz}(q_x,\omega)$ for $(\alpha,D/J)=(0,-0.2)$ at (a), (b) $h_s=0$ and (c), (d) $h_s/J=0.1$.
The excitation energies of the quantized excitation spectra of (e) $S^{xx}(q_x,\omega)$ and (f) $S^{zz}(q_x,\omega)$ at $q_x=\pi$ and $\pi/2$.
The constants used for the fitting are $\gamma_1 \approx 0.127$, $\gamma_2 \approx 0.155$, $\gamma_3 \approx 0.163$, and $\gamma_4 \approx 0.170$.}
\end{center}
\end{figure}
In Fig. \ref{fig3}, the DSFs for $(\alpha,D/J)=(0,-0.2)$ are shown, where the ground state is in the Haldane phase when $h_s=0$. 
For $h_s=0$, the single-magnon isolated mode appears in the low-lying excitation at $q_x \approx \pi$ and the multi-magnon continuum appears above the isolated mode,
which is schematically the same behavior as that of the isotropic case $D=0$\cite{SRWhite2,IAffleck,SYamamoto,MDPHorton,MTakahashi,FHLEssler}. 
Even for $(\alpha,D/J)=(0,-0.2)$, the quantization of the excitation continuum emerges for $h_s>0$.
However, the quantization is not clear in comparison with that for $D/J=-5$. 
In Figs. \ref{fig3}(e) and \ref{fig3}(f), the NZAF and excitation energies of the quantized spectra at $q_x=\pi$ and $\pi/2$, respectively, are compared.
For $(\alpha,D/J)=(0,-0.2)$, the excitation energies of the quantized spectra in $S^{xx}(q_x,\omega)$ and $S^{zz}(q_x,\omega)$ deviate from NZAF.
Note that the energy of the third quantized state (N=3) in $S^{zz}(q_x$=$\pi,\omega)$ accidentally agrees with that of NZAF.
Therefore, the origin of quantization in the excitation spectra is considered to be different from that in the Ising limit.

\begin{figure}[h]
\begin{center}
\includegraphics[bb=0 0 2240 1540, width=\linewidth]{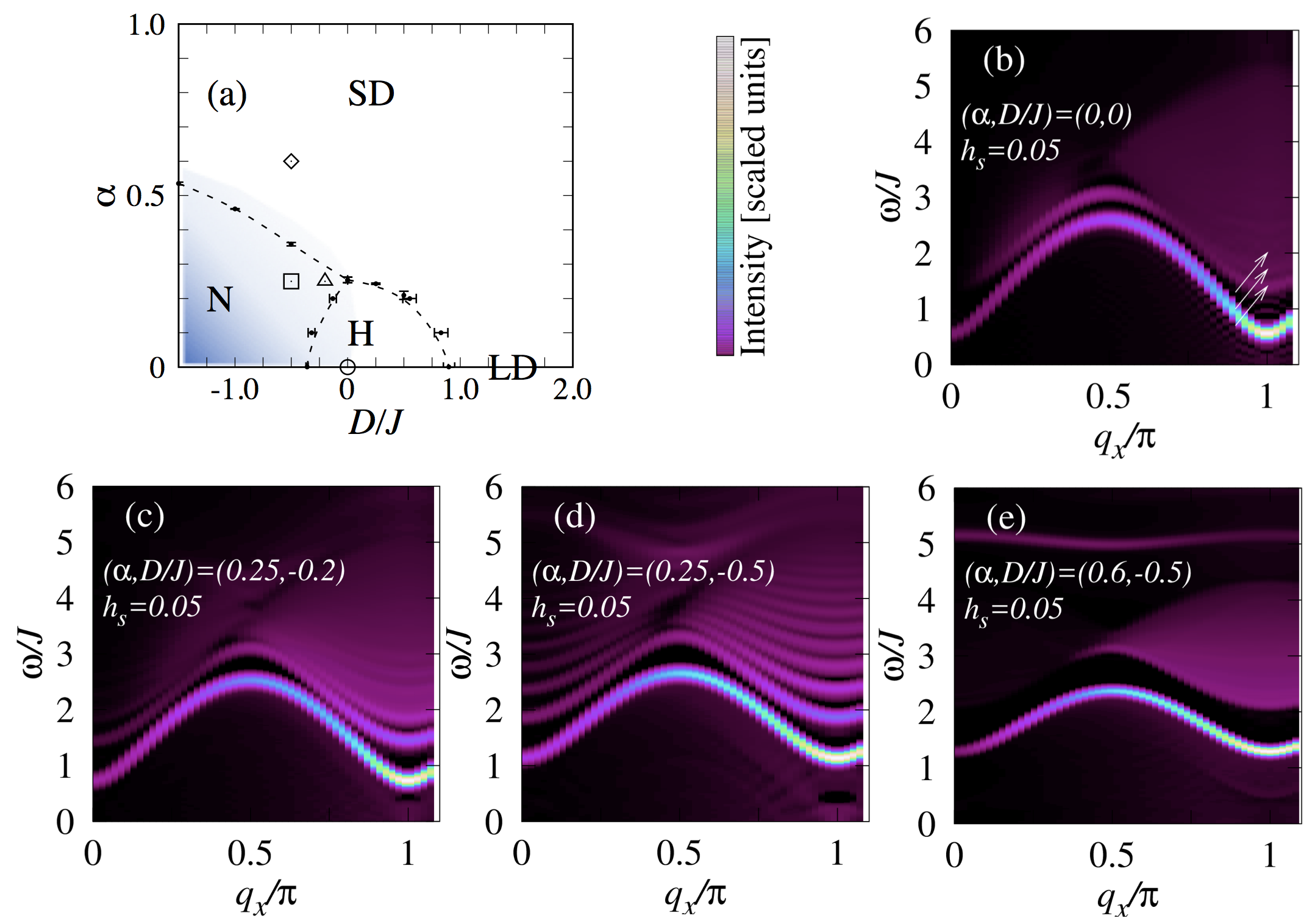}
\hspace{0pc}
\vspace{-1.5pc}
\caption{\label{fig4} (a) Schematic region where the quantized excitation spectra are observed. 
In the shaded area, the excitation continuum shows the quantization for $h_s>0$.
Black dots are critical points at $h_s=0$, presented in Ref. \onlinecite{TTonegawa}. 
"H," "N," "SD," and "LD" correspond to the Haldane phase, N\'eel phase, singlet-dimer phase, and large-$D$ phase, respectively.
(b)--(e) $S^{xx}(q_x,\omega)$ at $(\alpha,D/J)=(0,0)$, $(0.25,-0.2)$, $(0.25,-0.5)$, and $(0.6,-0.5)$ for $h_s=0.05$.
Open circle, triangle, square, and diamond in (a) correspond to the parameter used in (b)--(e) for $h_s=0$, respectively.  
Note that the DSFs are shown in the extended zone representation except for (b).  
White arrows in (b) indicate the peak positions of the intensity}
\end{center}
\end{figure}

For $(\alpha,D/J)=(0,-0.2)$ and $h_s=0$, the system is in proximity to the phase boundary between the Haldane and N\'eel phases.
At the critical point, the energy gap closes and the system is described by the Tomonaga-Luttinger liquid in the low-energy limit\cite{JTamaki}, which is equivalent to the theory of free boson fields.  
When the system departs from the critical point, the low-energy effective model is expressed by adding the family of $C_b \exp[i b \phi]$ terms\cite{TEXT} to the Lagrangian of free-boson field theory, where $\phi$ is the bosonic field and $b$ and $C_b$ are constants.
Each $C_{b} \exp[i b \phi]$ is classified as relevant or irrelevant, and the most relevant interaction opens the energy gap.
This means that the sine-Gordon field theory qualitatively describes the low-energy part of the system in proximity to the phase boundary.
In the excitation spectrum of the sine-Gordon field theory,  several isolated modes originating from the soliton/anti-soliton and breather modes are present in addition to the excitation continuum~\cite{JTamaki}.
When the system approaches the critical point, the excitation continuum shifts to the lower energy region and the isolated modes become unstable by touching the lower edge of the excitation continuum.
However, when the system deviates from the critical point, these isolated modes are placed below the lower edge of the excitation continuum and thus, we observe the quantized excitation spectra. 
This scenario for the quantized excitation spectra is considered valid in the vicinity of the following two critical lines:
One is dividing the N\'eel and Haldane phases, and the other is dividing the N\'eel and singlet-dimer phases, as shown in Fig. \ref{fig4}(a).

In Fig. \ref{fig4}(a), the region where the quantized excitation spectra appear for $h_s > 0$ is schematically represented by the shaded area.
Figures \ref{fig4}(b)--\ref{fig4}(e) are the results of $S^{xx}(q_x,\omega)$ at $(\alpha,D/J)=(0,0)$, $(0.25,-0.2)$, $(0.25,-0.5)$, and $(0.6,-0.5)$.
Since the large intensity appears at $q_x=\pi$ and $\omega=0$ in $S^{zz}(q_x,\omega)$ deep in the N\'eel phase\cite{SM},
the relative intensity of the quantized excitation spectra in $S^{zz}(q_x,\omega)$ is very weak.

For $(\alpha,D/J)=(0,0)$, $(0.25,-0.2)$, and $(0.25,-0.5)$, the system with the first parameter is located in the Haldane phase and the latter two are located in the N\'eel phase for $h_s=0$.
In such a case, the excitation continuum in the DSF is quantized for $h_s>0$, as shown in Figs. \ref{fig4}(b)--\ref{fig4}(d).
The quantization of the spectra is smeared when the system approaches the phase boundary in the singlet-dimer phase. 
The quantized excitation spectra disappear in the singlet-dimer phase, as shown in Fig. \ref{fig4} (e). 
This is explained by considering the low-energy excitation for $D=0$ and $\alpha \approx 1$.
For $D=0$ and $\alpha \approx 1$, the ground state is the direct product state of the singlet dimers.
The low-energy excitation that has a potential to compose the excitation continuum is given by replacing two singlets with two triplets from the ground state.
As $\alpha$ decreases from $\alpha=1$, the two-triplet excitation composes the excitation continuum around $q_x=\pi$\cite{TSuzuki}. 
However, the staggered field does not work as the confinement potential and rather localizes the triplet dimers with $S^{z}=0$.
Note that the energy of the triplet dimer with $S^{z}=\pm 1$ remains unchanged by the staggered field, which means that the confinement potential is absent.

At $D \rightarrow \infty$, the ground state is expressed by the direct product of $S^z=0$ at each site, namely, $\psi_0 = | \cdots $00000000$ \cdots \rangle$.
The lowest-energy excitation from $\psi_0$ is given by the single-spin flipping, $\psi^{+}_1 = | \cdots $000+000$ \cdots \rangle$ or $\psi^{-}_1 = | \cdots $000$-$000$ \cdots \rangle$.
The site carrying $S^z=\pm 1$ propagates by the hopping term, $S^+_iS^-_{i+1}+S^{-}_{i}S^{+}_{i+1}$, and $\psi^{\pm}_1$ composes the isolated mode.
The candidate for the low-energy excitation that composes the excitation continuum is obtained by creating two $S^{z}=1$ ($S^{z}=-1$) states at the neighboring sites, $\psi^{+}_2  = | \cdots $000++000$ \cdots \rangle$ or $\psi^{-}_2 = | \cdots $000$-$$-$000$ \cdots \rangle$.
In $\psi^{\pm}_2$, each site with $S^{z}=1$ (or $S^{z}=-1$) almost freely travels in the chain, and thus, they compose the excitation continuum.
However, since all sites sandwiched by the two sites with $S^{z}=1$ (or $S^{z}=-1$) are filled by $S^z=0$ due to the energy cost, the staggered field does not work as the confinement potential for the two sites with $S^{z}=0$.
Thus, the quantized excitation spectra are suppressed in the singlet-dimer phase\cite{SM}.


So far, many S=1 chain materials have been synthesized.
Most of them are considered to have the positive single-ion anisotropy, but negative single-ion anisotropy has been also reported for several materials.
For example, Y$_2$BaNiO$_5$\cite{GXu} and SrNi$_2$V$_2$O$_8$\cite{AKBera} have been evaluated as $D/J \approx -0.033$  and $\approx -0.057$, respectively.
There are likely many compounds whose proper models have been unsettled precisely due to the large deviation from the ideal Haldane-gap system.
In such compounds, there is a chance to find the quantized excitation spectra.
Further experiments are desirable.

We thank R. Coldea, B. Grenier, and S. Takayoshi for fruitful discussions.  
This work was supported by the CDMSI, CBSM2, and KAKENHI (Grants No. 15K05232 and No. 16K17751) from MEXT, Japan. 
T.S. thanks the computational resources of the K computer provided by the RIKEN AICS through the HPCI System Research Project (hp170262, hp170263, hp180170, and hp180225). 
We are also grateful for the numerical resources at the ISSP Supercomputer Center at the University of Tokyo  
and the Research Center for Nano-Micro Structure Science and Engineering at University of Hyogo.


\clearpage
\newpage
\pagebreak

\onecolumngrid
\begin{center}
  \textbf{\large Supplemental Materials: Quantized excitation spectra by magnon confinement in quasi-one-dimensional $S=1$ spin systems}\\[.2cm]
  Takafumi Suzuki$^1$ and Sei-ichiro Suga$^1$\\[.1cm]
  {\itshape ${}^1$Graduate School of Engineering, University of Hyogo, Himeji 671-2280, Japan\\}
  ${}^*$Electronic address: takafumi-s@eng.u-hyogo.ac.jp\\
(Dated: \today)\\[1cm]
\end{center}
\twocolumngrid

\setcounter{equation}{0}
\setcounter{figure}{0}
\setcounter{table}{0}
\setcounter{page}{1}
\renewcommand{\theequation}{S\arabic{equation}}
\renewcommand{\thefigure}{S\arabic{figure}}
\renewcommand{\bibnumfmt}[1]{[S#1]}
\renewcommand{\citenumfont}[1]{S#1}

\section*{\label{sec:level0} Benchmark test of dynamical spin structure factors by the iTEBD method}
\begin{figure}[htb]
\begin{center}
\includegraphics[bb=0 0 1350 1840, width=75mm]{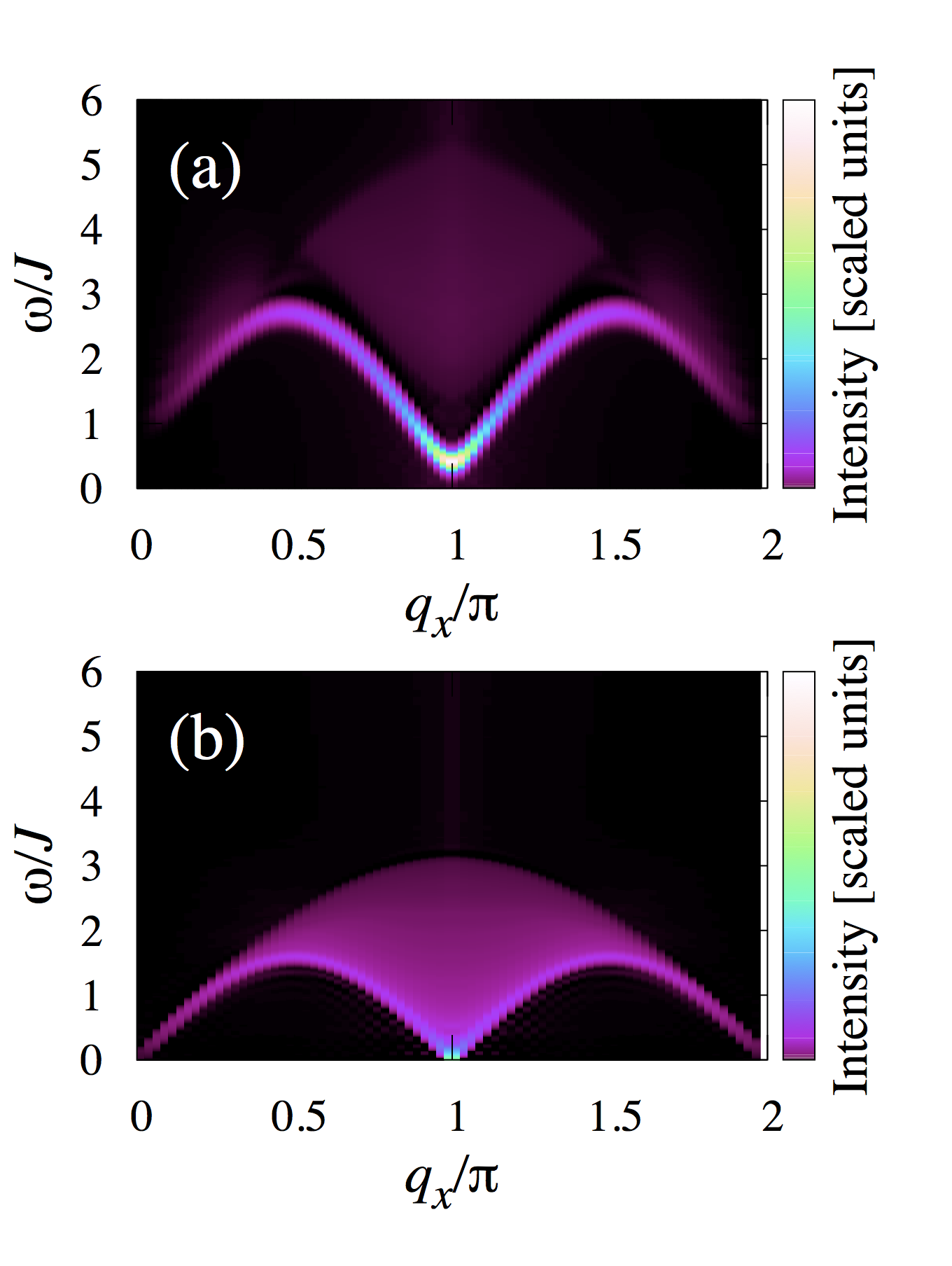}
\hspace{0pc}
\vspace{-1pc}
\caption{\label{figs1} Dynamical spin-structure factor.
(a) The $S=1$ AF Heisenberg chain. (b) The $S=1/2$ AF Heisenberg chain.} 
\end{center}
\end{figure}

The infinite time-evolving block decimation (iTEBD) algorithm\cite{GVidal,ROrus} is applied to calculate the dynamical spin structure factor (DSF)\cite{HNPhien,SRWhite}. 
The details of the numerical techniques have been discussed in Ref. \onlinecite{HNPhien}.
In Fig. \ref{figs1}, the benchmark results for the DSFs of the $S=1$ and $S=1/2$ antiferromagnetic (AF) Heisenberg chains are shown. 
In the computations, we set $\chi_{\rm max}=80$, which is the maximum bond dimension of the matrix-product-state representation for the wave function, and $N=200$ for the finite-size window\cite{HNPhien}.

In the $S=1$ AF Heisenberg chain, the result succeeds in reproducing the low-lying excitation of the Haldane-gap state, which has been already demonstrated in the previous works\cite{SRWhite,SYamamoto,HNPhien}.   
The isolated mode by a single magnon appears with the energy gap $\omega_{\rm H}/J \sim 0.41$ at $q_x = \pi$\cite{SRWhite,SYamamoto,HNPhien}
and the lower edge of the two-magnon continuum appears with the energy gap $\omega \sim 2 \omega_{\rm H}$ at $q_x \approx 0$~\cite{IAffleck}.
In the low-lying excitation, the isolated mode is stable for $0.3 \lessapprox q_x \le \pi$ and becomes unstable for $0 \le q_x \lessapprox 0.3\pi$\cite{SYamamoto,HNPhien}. 
Above the single-magnon mode, the excitation continuum appears at $q_x \approx \pi$. 
The lower edge of the continuum is composed by the three-magnon excitation.\cite{MDPHorton}

For the $S=1/2$ AF Heisenberg chain, we reproduce the dispersion relation given by des Cloizeaux and Pearson~\cite{JCLoizeqaux} and the upper boundary of the spinon continuum~\cite{TYamada,GMuller,MKarbach}.
Although it is difficult to confirm the singularity of the intensity at the lower edge and higher edge of the excitation continuum quantitatively,
the well-known spinon continuum is reproduced qualitatively ~\cite{JCLoizeqaux,TYamada,GMuller,MKarbach,KAHallberg}. \\

\section*{\label{sec:level1} Quantized spectra and negative zeros of the Airy function}
McCoy and Wu argued the relationship between the negative zeros of the Airy function (NZAF) and the confinement of the spinon excitations 
from the view point of the excitation spectra of the $S=1/2$ ferromagnetic Ising chain with transverse fields and weak longitudinal magnetic fields. 
When we start from the ferromagnetic ordered state, the low-energy excitation is achieved by flipping all spins in a single domain with arbitrary length. 
Since each domain carries $\Delta S=1/2$, namely a single spion, the lowest excitation is described by the two-spion state. 
When the transverse field exists, the two spinons travel on the chain and compose an excitation continuum. 
In addition, the uniform longitudinal field is further applied weakly, and then the linear potential works between the two spinons.
The effective model for the kinetics of the two spinons is written as
\begin{eqnarray}
\left[ \frac{\partial^2}{\partial x^2} + c|x| \right] \phi = \epsilon \phi,
\label{eq1}
\end{eqnarray} 
where $c$ is a positive constant. 
The eigenvalues of Eq. (\ref{eq1}) are exactly solved and given by NZAF. 
This means that the two-spinon continuum is quantized by the weak uniform longitudinal field. 
Shiba also discussed the quantized excitation spectra in the quasi-one-dimensional $S=1/2$ antiferromagnetic (AF) Ising-like XXZ chain\cite{HShiba}. 
When we focus on the low-energy excitation in the bipartite systems, 
the inter-chain interaction works as the staggered field in spin chains.
After $\pi$ rotations for the $s^x$ axis on one of the two sub-lattices,
the staggered field becomes equivalent to the longitudinal uniform field.
Thus, the low-energy excitation of the the same kinetics of the domain-wall excitation is obtained in the AF case.

\section*{\label{sec:level1} Quantized spectra in the $S=1$ bond-alternating spin chain}
\begin{figure*}[h]
\begin{center}
\includegraphics[bb=0 0 1990 1990, width=160mm]{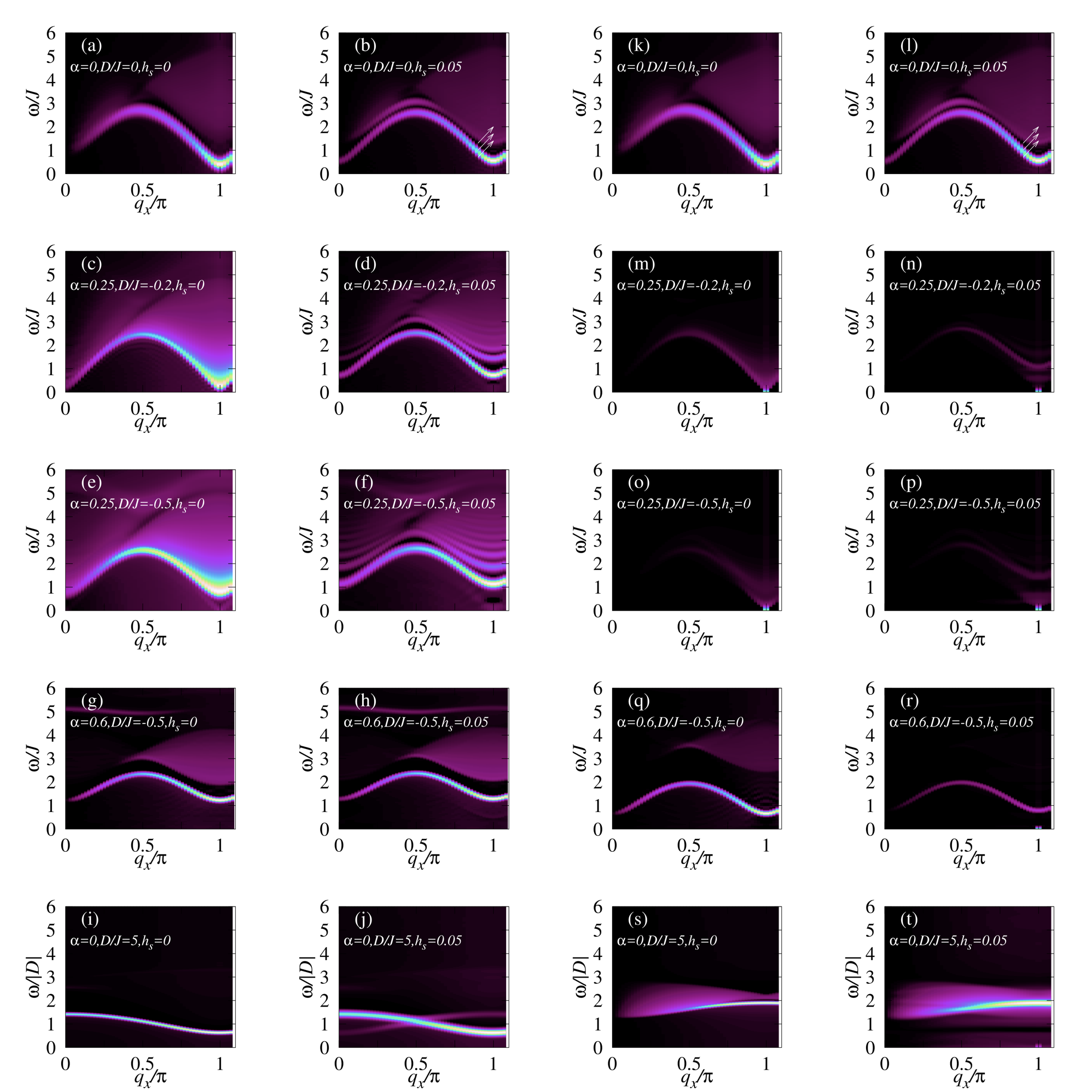}
\hspace{0pc}
\vspace{0pc}
\caption{\label{figs2}(a)--(j) $S^{xx}(q_x,\omega)$. (k)--(t)$S^{zz}(q_x,\omega)$. 
From the top row to the bottom row, the results for $(\alpha,D/J)=(0,0)$, $(0.25,-0.2)$, $(0.25,-0.5)$, $(0.6,-0.5)$, and $(0,5)$ are presented, respectively.
The leftmost column (second column) and the third column (rightmost column) are the results for $h_s=0$ ($h_s>0$). 
White arrows are pointing peak positions.}
\end{center}
\end{figure*}
In the main text,  we focused on the results of $S^{xx}(q_x,\omega)$ for $h_s>0$.
In this supplemental material, the results of $S^{xx}(q_x,\omega)$ and $S^{zz}(q_x,\omega)$ for $h_s=0$ and $h_s>0$ are shown.
Figures \ref{figs1}(a)--\ref{figs1}(t) are the results for $(\alpha,D/J)=(0,0)$, $(0.25,-0.2)$, $(0.25,-0.5)$, $(0.6,-0.5)$, and $(0,5)$.
The definitions of $\alpha$, $D$, and $h_s$, are presented in the main text.
When the system is located in the Haldane phase with $D \lessapprox 0$, 
the quantized excitation spectra appear in $S^{xx}({q_x},\omega)$ and $S^{zz}(q_x,\omega)$ [Figs. \ref{figs2}(b) and \ref{figs2}(l)].
The quantized excitation spectra in $S^{xx}(q_x,\omega)$ become clear for $h_s>0$, as the system moves to the deep N\'eel phase [Figs. \ref{figs2}(d) and \ref{figs2}(f)].
Although, in the deep N\'eel phase, $S^{zz}(q_x,\omega)$ shows the quantized excitation spectra, the quantization is not clear 
because the peak at $\omega \approx 0$ shows the quite large intensity [Figs. \ref{figs2}(n) and \ref{figs2}(p)].
When the system is in the singlet-dimer phase, the staggered field does not contribute the confinement of the domain-wall excitation.
Thus, the quantized excitation spectra do not appear, as shown in Figs. \ref{figs2}(h), \ref{figs2}(j), \ref{figs2}(r), and \ref{figs2}(t).

\end{document}